\documentclass[longauth]{aa} 
\usepackage{graphicx}
\usepackage{tabularx}
\usepackage{multirow}
\usepackage[colorlinks=true, linkcolor=blue, citecolor=blue]{hyperref}
\usepackage{gensymb}
\usepackage{color}
\usepackage{txfonts}
\usepackage{natbib}
\usepackage{amsmath}
\usepackage{ulem}
\usepackage[flushleft]{threeparttable}
\usepackage[dvipsnames]{xcolor}
\usepackage{rotating}
\usepackage{CJKutf8}

\begin{document} 
   \title{PANORAMIC: Discovery of an Ultra-Massive Grand-Design Spiral Galaxy at $z\sim5.2$}

\author{ Mengyuan Xiao\inst{\ref{inst1}}\thanks{E-mail: mengyuan.xiao@unige.ch}
\and Christina C. Williams\inst{\ref{inst2},\ref{inst3}} 
\and Pascal A. Oesch\inst{\ref{inst1},\ref{inst4}} 
\and David Elbaz\inst{\ref{inst5}} 
\and Miroslava Dessauges-Zavadsky\inst{\ref{inst1}} 
\and Rui Marques-Chaves\inst{\ref{inst1}}  
\and Longji Bing\inst{\ref{inst6}} 
\and Zhiyuan Ji\inst{\ref{inst3}} 
\and Andrea Weibel\inst{\ref{inst1}} 
\and Rachel Bezanson\inst{\ref{inst7}} 
\and Gabriel Brammer\inst{\ref{inst4}} 
\and Caitlin Casey \inst{\ref{inst8},\ref{inst9},\ref{inst4}} 
\and Aidan P. Cloonan \inst{\ref{inst10}} 
\and Emanuele Daddi\inst{\ref{inst5}} 
\and Pratika Dayal\inst{\ref{inst11}}
\and Andreas L. Faisst\inst{\ref{inst12}}
\and Marijn Franx\inst{\ref{inst13}}
\and Karl Glazebrook\inst{\ref{inst14}}
\and Anne Hutter\inst{\ref{inst4}}
\and Jeyhan S. Kartaltepe\inst{\ref{inst15}}
\and Ivo Labbe\inst{\ref{inst14}}
\and Guilaine Lagache\inst{\ref{inst16}}
\and Seunghwan Lim\inst{\ref{inst17},\ref{inst18}} 
\and Benjamin Magnelli\inst{\ref{inst5}} 
\and Felix Martinez\inst{\ref{inst15}}
\and Michael V. Maseda \inst{\ref{inst19}}
\and Themiya Nanayakkara\inst{\ref{inst14}}
\and Daniel Schaerer\inst{\ref{inst1}}
\and Katherine E. Whitaker \inst{\ref{inst10}} 
}

\institute{Department of Astronomy, University of Geneva, Chemin Pegasi 51, 1290 Versoix, Switzerland\label{inst1}
\and NSF National Optical-Infrared Astronomy Research Laboratory, 950 North Cherry Avenue, Tucson, AZ 85719, USA \label{inst2}
\and Steward Observatory, University of Arizona, 933 North Cherry Avenue, Tucson, AZ 85721, USA \label{inst3}
\and  Cosmic Dawn Center (DAWN), Niels Bohr Institute, University of Copenhagen, Jagtvej 128, K\o benhavn N, DK-2200, Denmark \label{inst4}
\and Universit{\'e} Paris-Saclay, Universit{\'e} Paris Cit{\'e}, CEA, CNRS, AIM, 91191, Gif-sur-Yvette, France \label{inst5}
\and Astronomy Centre, University of Sussex, Falmer, Brighton BN1 9QH, UK  \label{inst6}
\and Department of Physics and Astronomy and PITT PACC, University of Pittsburgh, Pittsburgh, PA 15260, USA \label{inst7}
\and Department of Physics, University of California, Santa Barbara, CA 93106, USA\label{inst8}
\and Department of Astronomy, University of Texas at Austin, 2515 Speedway Blvd. Stop C1400, Austin, TX 78712, USA\label{inst9}
\and Department of Astronomy, University of Massachusetts Amherst, Amherst MA 01003, USA\label{inst10}
\and Kapteyn Astronomical Institute, University of Groningen, PO Box 800, 9700 AV Groningen, The Netherlands \label{inst11}
\and Caltech/IPAC, 1200 E. California Blvd. Pasadena, CA 91125, USA \label{inst12}
\and Leiden Observatory, Leiden University, P.O.Box 9513, NL-2300 AA Leiden, The Netherlands \label{inst13}
\and Centre for Astrophysics and Supercomputing, Swinburne University of Technology, P.O. Box 218, Hawthorn, 3122, VIC, Australia \label{inst14}
\and Laboratory for Multiwavelength Astrophysics, School of Physics and Astronomy, Rochester Institute of Technology, 84 Lomb Memorial Drive, Rochester, NY 14623, USA \label{inst15}
\and Aix Marseille Univ, CNRS, CNES, LAM, Marseille, France \label{inst16}
\and Kavli Institute for Cosmology, University of Cambridge, Madingley Road, Cambridge, CB3 0HA, UK \label{inst17}
\and Cavendish Laboratory, University of Cambridge, 19 JJ Thomson Avenue, Cambridge, CB3 0HE, UK\label{inst18}
\and Department of Astronomy, University of Wisconsin-Madison, 475 N. Charter St., Madison, WI 53706, USA\label{inst19}
}
\date{Received December 17, 2024; accepted February 27, 2025}
 
  \abstract
 {We report the discovery of an ultra-massive grand-design red spiral galaxy, named Zhúlóng (Torch Dragon), at $z_{\rm phot} = 5.2^{+0.3}_{-0.2}$ in the JWST PANORAMIC survey, identified as the most distant bulge+disk galaxy candidate with spiral arms known to date. Zhúlóng displays an extraordinary combination of properties:  1) a classical bulge centered in a large, face-on exponential stellar disk (half-light radius of $R_{\rm e} = 3.7 \pm 0.1 \, \mathrm{kpc}$), with spiral arms extending across 19 kpc;  2) a clear transition from the red, quiescent core ($F150W-F444W=3.1$ mag) with high stellar mass surface density ($\log(\Sigma M_{\star}/M_{\odot} \, \mathrm{kpc}^{-2}) = 9.91_{-0.09}^{+0.11}$) to the star-forming outer regions, as revealed by spatially resolved SED analysis, which indicates significant inside-out galaxy growth; 3) an extremely high stellar mass at its redshift, with $\log (M_{\star}/M_{\odot})=11.03_{-0.08}^{+0.10}$ comparable to the Milky Way, and an implied baryon-to-star conversion efficiency ($\epsilon \sim 0.3$) that is 1.5 times higher than even the most efficient galaxies at later epochs; 4) despite an active disk, a relatively modest overall star formation rate ($\mathrm{SFR} =66_{-46}^{+89} ~M_{\odot} \, \mathrm{yr}^{-1}$), which is $>$0.5 dex below the star formation main sequence at $z \sim 5.2$ and $>$10 times lower than ultra-massive dusty galaxies at $z=5-6$. Altogether, Zhúlóng shows that mature galaxies emerged much earlier than expected in the first billion years after the Big Bang through rapid galaxy formation and morphological evolution. 
Our finding offers key constraints for models of massive galaxy formation and the origin of spiral structures in the early universe.  }

   \maketitle

\section{Introduction}

A remarkable revelation from early James Webb Space Telescope (JWST) observations is the rapid evolution of galaxies at early cosmic times \citep[see review in][]{Adamo2024}. Three years into the mission \citep{Gardner2023}, evidence for faster growth and maturity among galaxies continues to accumulate. By nearly all common evolutionary metrics (stellar mass, luminosity, stellar structure), galaxies have evolved faster than expected prior to JWST \citep[e.g.,][]{Naidu2022,  Boylan-Kolchin2023, Donnan2023, Ferreira2023, Finkelstein2024, Casey2024, deGraaff2024, Glazebrook2024, Shapley2024}. This raises questions about the physical processes that drive the inferred rapid timescales associated with the observed evolution, and/or points to an earlier onset of galaxy formation. 

At the forefront of these discoveries are massive galaxies. Panchromatic JWST imaging and spectroscopy from the observed optical to mid-infrared have confirmed not only that ultra-massive galaxies existed in the early Universe ($\log (M_{\star}/M_{\odot})>11$ at $z>5$), but that they may also exist at unexpectedly high abundance \citep[e.g.][]{Labbe2023a, Xiao2024, Weibel2024}. They are so massive, with almost Milky Way (MW) mass only at $\sim$1~Gyr after the Big Bang, suggesting that galaxies are forming in a more efficient way than previously thought. Based on their existence in small survey areas (typical JWST imaging campaigns from Cycle 1 $<100$ sq arcmin per field), these ultra-massive galaxies, like the ones reported in \cite{Xiao2024}, require on average about 50\% of the baryons in their dark matter halos to be converted into stars – two to three times more than even the most efficient galaxies at later times \citep{Moster2013,Wechsler2018,Shuntov2022}. While questions remain about incomplete data or modeling assumptions for photometric samples \citep[e.g.][]{Wangbingjie2024,Williams2024_RED,Desprez2024}, cosmic variance from small fields and sample sizes also complicate interpretation. This motivates studies of known massive candidates, because if true, these findings definitively raise new important questions about the theory of evolutionary drivers of early galaxies. 

A key constraint on evolutionary drivers of rapid galaxy growth is galaxy morphology. On this front, JWST is again producing surprising revelations pointing to accelerated maturity timescales for galaxies, when compared to the morphological census using the Hubble Space Telescope (HST). Prior to JWST, HST imaging suggested the emergence of disks and spiral galaxies (perhaps the pre-cursors of grand-design spirals seen locally) was a late forming phenomenon, with the backbone of the Hubble Sequence appearing as late as $z\sim2$ \citep[e.g.][]{BLee2013, Mortlock2013}. However, with JWST, there is now accumulating evidence that stellar disks formed earlier than we expected, making up $\sim50\%$ of galaxies up to $z<6$ \citep{Ferreira2022,Ferreira2023, VegaFerrero2024, Robertson2023, Kartaltepe2023, Jacobs2023,HuertasCompany2024}. This is a factor of 10$\times$ more stellar disks than were thought to exist at this redshift prior to JWST. In addition, a few stellar spiral galaxies have also been reported at $z>3$ \citep{jain2024, Wangweichen2024, Umehata2024, Costantin2023, Wu2023}. These new findings contrast with theoretical predictions that large stellar disks should predominantly form later in cosmic history \citep[e.g.,][]{Dekel2020,Belokurov2022, Gurvich2023,McCluskey2024}. 

Morphological constraints are of particular interest regarding ultra-massive galaxies, where the stellar structure lends important clues to the highly efficient buildup that is implied by their high masses at early times. To date, numerous extremely compact, massive quiescent galaxies have been spectroscopically confirmed with stellar absorption features at $z\gtrsim5$, and are extremely compact \citep[e.g.,][]{Carnall2023Natur, Weibel2024b, deGraaff2024}. These findings mirror the established compact sizes of quiescent galaxies at cosmic noon, which are typically a factor of $>3$ smaller in size at fixed mass compared to star-forming galaxies \citep[e.g.,][]{vandokkum2008,vanderwel2014, faisst2017, Kawinwanichakij2021, Mowla2019b, Cutler2022}. However, their compact structures are not a universal feature: three extreme star-forming examples of ultra-massive galaxies presented in \citet{Xiao2024} exhibit larger and irregular morphologies, raising the question of whether dust obfuscates the stellar morphology or if there is diversity in the population. With relatively few ultra-massive galaxies known (identified from relatively small NIRCam blank-field areas), a systematic study to characterize the morphologies of representative samples has not yet been carried out, preventing general conclusions about formation pathways using morphology.

Here, we report on the serendipitous discovery of a unique ultra-massive galaxy at $z_\mathrm{phot}\sim5.2$ in the PANORAMIC\footnote{Parallel wide-Area Nircam Observations to Reveal And Measure the Invisible Cosmos (PANORAMIC)} Survey \citep{Williams2024}. Unlike the other ultra-massive candidates described above, it has a striking evolved morphology: a quiescent-like classical bulge + star-forming stellar disk + grand-design spiral arms (defined as spiral structures spanning the whole galaxy, from the nucleus to outskirts, with spiral arms starting at diametrically opposite points; e.g. \citealt{Binney2008, Dobbs2014, Sellwood2022}), already at 1 billion years after the Big Bang. More interestingly, it is very large in stellar size (19 kpc in diameter), extremely massive ($\log (M_{\star}/M_{\odot})>11$, indicating high baryon-to-star formation efficiency), with a modest star formation rate with red color ($\mathrm{SFR} =66_{-46}^{+89} ~M_{\odot} \, \mathrm{yr}^{-1}$; F150W$-$F444W$\,\sim2.7$ mag). Considering only a few stellar spiral galaxies have been identified at $z>3$ \citep{jain2024, Wangweichen2024, Umehata2024, Costantin2023, Wu2023}, this source becomes the most distant massive spiral galaxies discovered so far, providing direct empirical evidence of the rapid dynamical evolution of massive galaxies at $z>5$. 
Given its unique red color, morphology, and stellar properties, we name it Zhúlóng\footnote{Zhúlóng (\begin{CJK}{UTF8}{gkai}烛龙\end{CJK}, or Torch Dragon), which is a giant red solar dragon and god in Chinese mythology. It supposedly had a human's face and snake's body, created day and night by opening and closing its eyes, and created seasonal winds by breathing.}.

This paper is organized as follows: In Section \ref{Sec: 2} we present the datasets used to measure photometry and photometric redshift methods. In Section \ref{Sec: measure} we present our methodology for measuring morphologies and SED modeling. In Section \ref{Sec: disc} we interpret the properties of this remarkable galaxy in the context of both the formation pathway of ultra-massive galaxies, and as the highest redshift example of a large, spiral galaxy.
Throughout this paper, we adopt a Chabrier initial mass function \citep[IMF;][]{Chabrier2003} to estimate SFR and stellar mass. We assume Planck cosmology \citep{Planck2020}. with $(\Omega_\mathrm{m},\, \Omega_\mathrm{\Lambda},\, h,\, \sigma_\mathrm{8})=(0.3,\, 0.7,\, 0.7, \,0.81)$.
When necessary, data from the literature have been converted with a conversion factor of $M_{\star}$ \citep[][IMF]{Salpeter1955} = 1.7  $\times$ $M_{\star}$ \citep[][IMF]{Chabrier2003}. All magnitudes are in the AB system \citep{Oke1983}, such that $m_{\rm AB} = 23.9 - 2.5$ $\times$ log(S$_{\nu}$ [$\mu$Jy]). 

\section{Observational data}\label{Sec: 2}

\noindent Zhúlóng ($\alpha$, $\delta$ [J2000] = 150.124874, 2.092919) was serendipitously discovered in the field (association) named j100024p0208 
in the recent data release of the JWST PANORAMIC survey \citep[GO-2514; PIs: C. Williams \& P. Oesch;][]{Williams2024}, a pure parallel extragalactic NIRCam imaging program. The source was originally identified by chance during a search for massive galaxy candidates that were color-selected to exhibit a strong Balmer break at $z>3$ \citep{Long2024}. Its spiral structure was identified during visual inspection of those candidates, and was unique among the sample.  
Zhúlóng lies within the NIRCam imaging footprint of the COSMOS-Web field \citep{Casey2023}, but is not covered by the COSMOS-PRIMER program \citep{PRIMER}.

\begin{figure*}
\centering
\includegraphics[width=17cm]{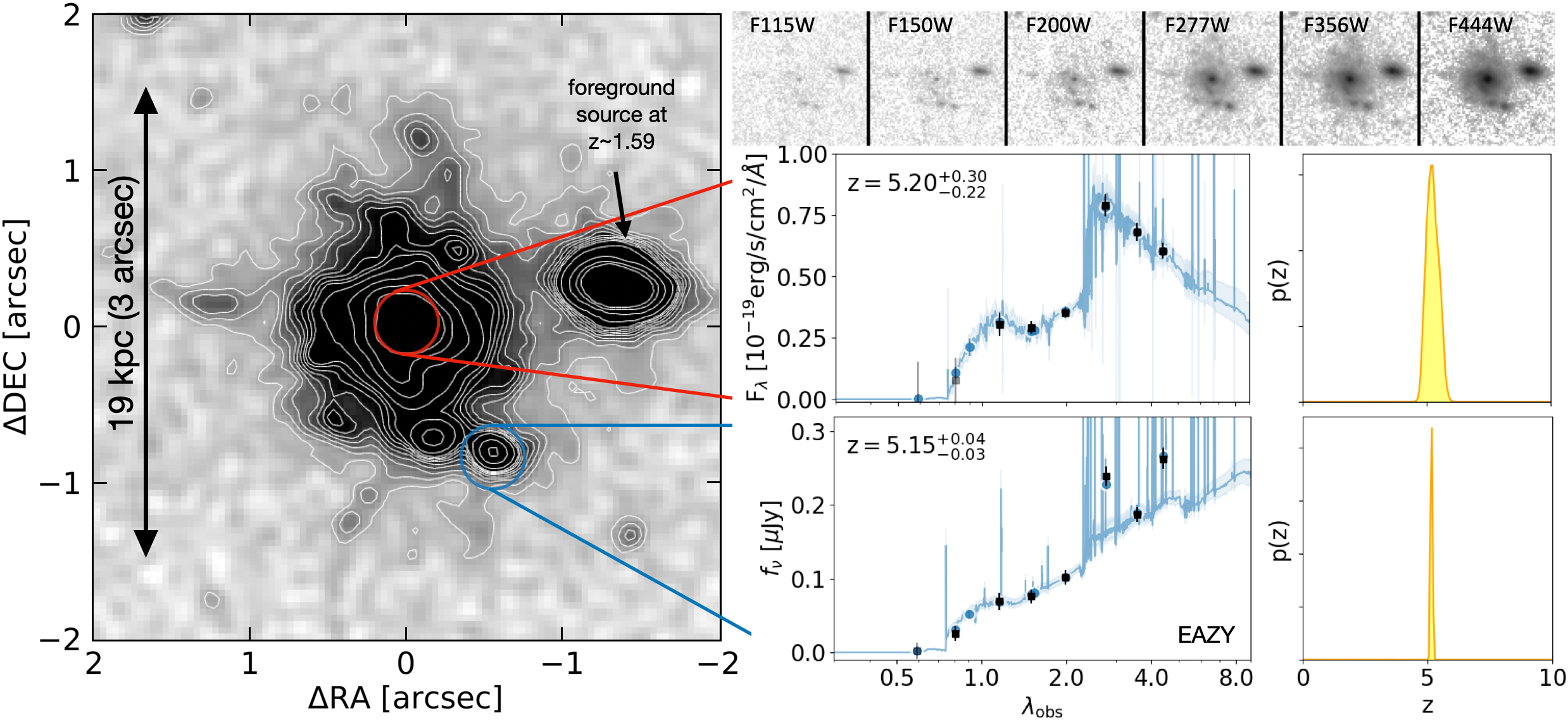}
\caption{\textbf{The morphology and photo-z of Zhúlóng}. The left panel shows a 4\arcsec$\times$4\arcsec stacked image (F277W+F356W+F444W) with contour maps (start from 3$\sigma$). The top right panels show these same stamps in different filters. The central core's SED (measured in 0.16$^{\prime\prime}$ radius; circled in red) has a clear post-starburst shape, which is shown in the right middle panels. The first panel shows the \texttt{EAZY} SED fit together with the photometric points, while the photometric redshift likelihood distribution is shown in the right panel, showing a single peak consistent with $z_{\rm phot} = 5.2^{+0.3}_{-0.2}$. A neighboring star-forming clump (circled in blue) shows a very robust redshift of $z_{\rm phot} = 5.15^{+0.04}_{-0.03}$ from photometry due to strong emission lines.
      }
         \label{fig1}
\end{figure*}

\subsection{HST and JWST NIRCam observations}\label{Sect:panoramic}
We utilize imaging data from archival HST observations, including F606W and F814W from HST/ACS \citep[][]{koekemoer2007,koekemoer2011,grogin2011}, and F160W from HST/WFC3  \citep[][]{Momcheva2017,Mowla2019}. The spiral galaxy is not detected in any HST images, and was discovered in JWST/NIRCam \citep{Rieke2023} images with filters F115W, F150W, F200W, F277W, F356W, and F444W obtained as part of the PANORAMIC Survey. For the four filters obtained by COSMOS-Web (F115W, F150W, F277W, F444W), we combine the imaging from the two surveys. The typical 5$\sigma$ depths, measured within a circular aperture of 0.16$^{\prime\prime}$ radius, are 27.5, 27.7, 28.2, 28.4, 28.8, and 28.3 mags for F115W, F150W, F200W, F277W, F356W, and F444W, respectively. 

Multi-wavelength photometric measurements are derived following the same procedure as outlined in \citet{Weibel2024}. Briefly, we use \texttt{SExtractor} \citep{Bertin1996} in dual image mode with an inverse-variance weighted stack of the longest wavelength NIRCam wide filters, F277W+F356W+F444W, as the detection image. In this study, fluxes are measured in 0\farcs16, 0\farcs35, 0\farcs5, and 0\farcs7 radius circular apertures in images that are point spread function (PSF)-matched to the F444W band. Total fluxes are derived from the Kron AUTO aperture provided by \texttt{SExtractor} in the F444W band, in addition to a correction based on the encircled energy of the Kron aperture on the F444W PSF. Detailed descriptions of data reduction and photometric measurements are provided in \cite{Williams2024}.

\subsection{Longer-wavelength observations}\label{alma}
Although Zhúlóng lies within the COSMOS-Web field, it falls in a gap in the MIRI F770W footprints. Observations at longer wavelengths cover the location of Zhúlóng, but none return a detection, which include \textit{Spitzer} MIPS 24 \textmu m \citep{LeFloch2009}, \textit{Herschel} PACS \citep{Lutz2011} and SPIRE \citep{oliver2012}, NIKA2 Cosmological Legacy Survey at 1.2mm and 2.0mm \citep[rms of 0.3 and 0.09 mJy beam$^{-1}$, respectively;][Carvajal Bohorquez et al. in prep]{Bing2023}, ALMA 2.0mm (rms of 0.12 mJy beam$^{-1}$; Project ID: 2021.1.00705.S), ALMA 3.0mm (rms of 0.06 mJy beam$^{-1}$; Project ID: 2021.1.01005.S), and VLA 3 GHz \citep{Smolcic2017} and 1.4 GHz \citep{Schinnerer2010}. More recently, the ALMA large program CHAMPS (PI: A. Faisst; priv. comm.; Faisst et al. in prep) also covers Zhúlóng at 1.2 mm, but again it remains undetected. The local rms level is $\sim$0.2 mJy beam$^{-1}$ in the 1.2mm map of $0.959^{\prime\prime}\times 0.799^{\prime\prime}$ spatial resolution. The 3$\sigma$ upper limit at 1.2mm is 0.6 mJy, assuming the dust distribution is unresolved, indicating that Zhúlóng has low far-infrared luminosity. 

We note that, with the exception of the ALMA 1.2 mm observations, the remaining long-wavelength observations are too shallow to significantly constrain the far-infrared SED shape of Zhúlóng (see Sect. \ref{Sec: mass}). Therefore, in this work, we only use data points from HST, JWST, and ALMA 1.2mm.

\subsection{Zhúlóng with robust photometric redshift: $z_{\rm phot} = 5.2^{+0.3}_{-0.2}$}

The photometric redshift of Zhúlóng is estimated using the SED-fitting code \texttt{EAZY} \citep{Brammer2008}, with the \texttt{blue\_sfhz\_13} template set which imposes redshift-dependent star formation histories (SFHs), excluding SFHs that start earlier than the age of the Universe at a given redshift\footnote{\url{https://github.com/gbrammer/eazy-photoz/tree/master/templates/sfhz}}.
We apply an error floor of 5\% prior to running \texttt{eazy} to account for possible remaining systematic uncertainties in the photometry and to allow for more flexibility in the SED-fitting. The redshift range for fitting is set to $z \in (0.01, 20)$. Additional details on the photometric redshift modeling are presented in \cite{Williams2024}.

The \texttt{EAZY} best-fit results are shown in Fig.~\ref{fig1}. The redshift probability distribution $P(z)$ for Zhúlóng is strongly constrained by a prominent Balmer/4000 \AA~ break observed in the central region (0.16$^{\prime\prime}$ radius), resulting in narrow uncertainties with a single redshift solution $z_{\rm phot} = 5.2^{+0.3}_{-0.2}$. In addition, the neighboring south-west clump has evidence of strong emission lines, which boost the photometry in the F277W and F444W bands, leading to an even better constrained photometric redshift: $z_{\rm phot} = 5.15^{+0.04}_{-0.03}$.  This consistency illustrates the $z_{\rm phot}$ reliability of Zhúlóng. Further tests and a detailed discussion of the robustness of the redshift are presented in Appendix~\ref{low-z}. Nevertheless, we underline that this is still a photometric redshift. It will be important to confirm the redshift through spectroscopy in the future, such as with the upcoming COSMOS-3D grism observations (GO-5893, \citealt{COSMOS-3D}) or with NIRSpec spectra. 
\begin{figure*}
\centering
\includegraphics[width=18.3cm]{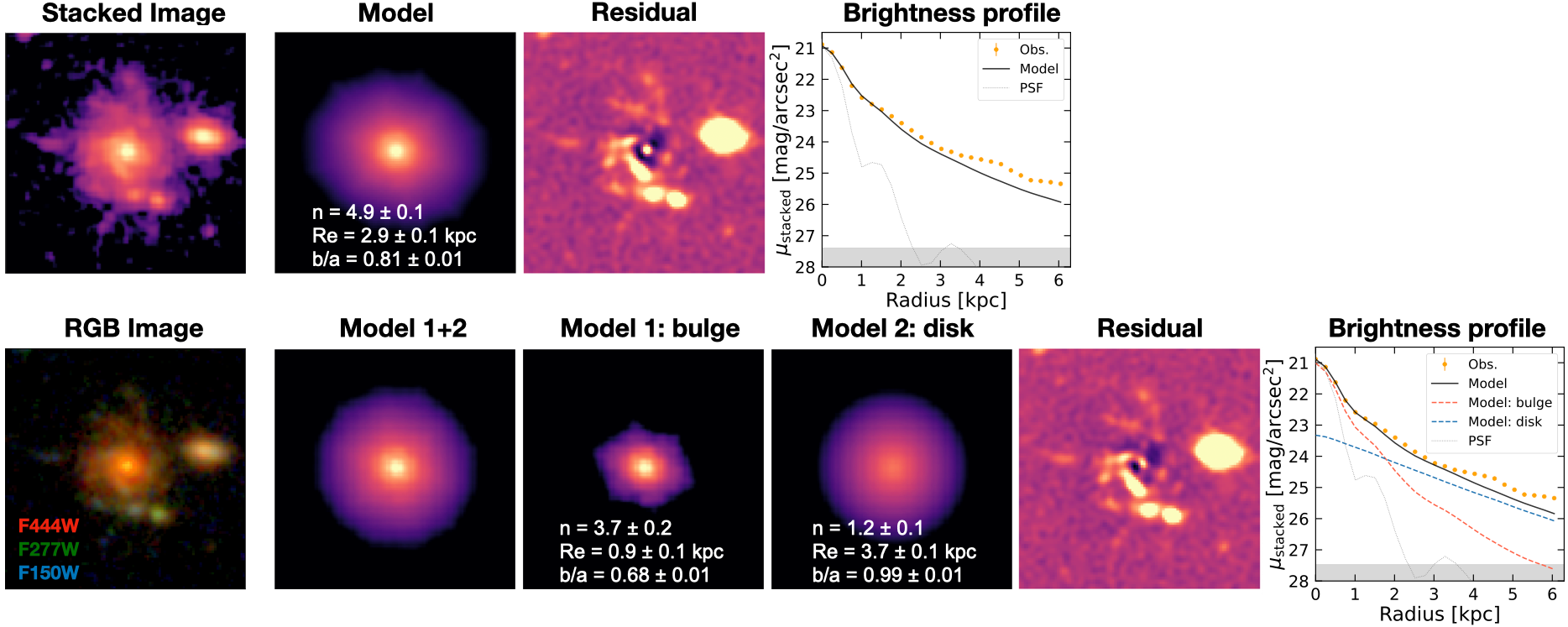}

\caption{\textbf{Morphological modeling of Zhúlóng.} The stacked image ($4^{\prime\prime}\times 4^{\prime\prime}$; F277W+F356W+F444W) is modeled using \texttt{PySersic} with: 1) a single S\'{e}rsic profile (top row) and 2) double S\'{e}rsic profiles (bottom row). Panels from left to right show: the stacked image, the best-fit model, the model-subtracted residuals, and the surface brightness profile. The grey-shaded area in the profile indicates the 1$\sigma$ noise level of the image. Notably, after subtracting the major component(s) of Zhúlóng, the spiral arms appear prominently in the residual maps. For the morphological analysis of individual images of F277W, F356W, and F444W, respectively, see Appendix~\ref{fit}.
      }
         \label{morphology_plot}
\end{figure*}

\begin{figure*}
\centering
\includegraphics[width=18.cm]{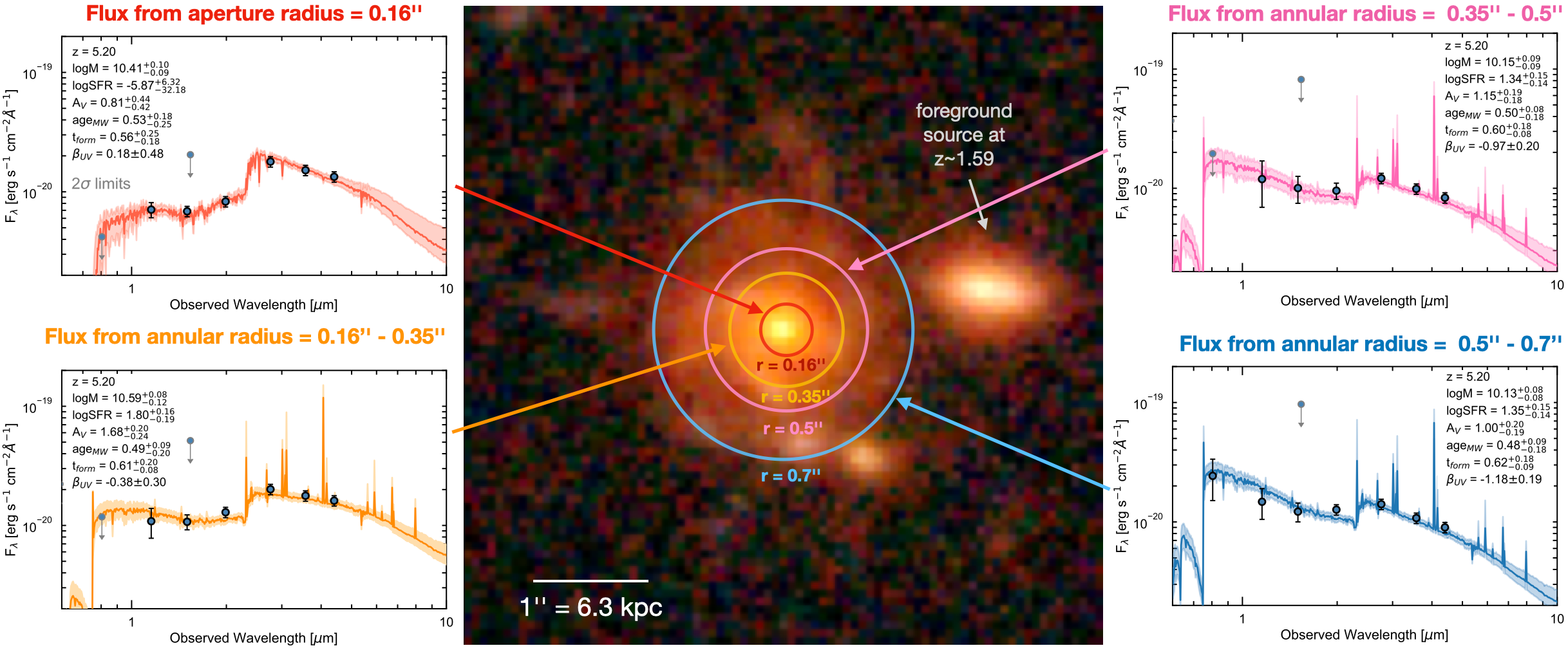}
\caption{\textbf{Stellar population gradient from core to outer region.} Zhúlóng has a quiescent-like galaxy core embedded in a star-forming stellar disk. The figure shows the SED shapes and photometry as measured in different annuli, from the central core (top left) to the outer region (bottom right), as indicated by the labels. The properties of different annular regions are listed in Table \ref{table1}.  Note that the galaxy is very extended, with spiral arms extending to more than 19 kpc in diameter. 
      }
         \label{fig3}
\end{figure*}

\begin{table*}
\caption{Properties of the entire galaxy and of the different annular regions.}   
\tiny          
\centering
\renewcommand{\arraystretch}{1.2} 
\begin{threeparttable} 
 
\begin{tabular}{l c c ccc }     
\hline\hline       
Total  & $z_{\rm phot}$ & log $M_{\rm \star} [M_{\odot}$] & \rm{SFR} [$M_{\odot}$yr$^{-1}$]  & $A_{\rm V}$ [mag] &  \\ 
\hline  
           &       $5.2^{+0.3}_{-0.2}$ &  11.03$_{-0.08}^{+0.10}$ &   66$_{-46}^{+89}$  &  1.1$_{-0.2}^{+0.2}$ &    \\
\hline
\\
\hline\hline     
Annulus radius  & log $\Sigma M_{\rm \star} [M_{\odot}$ kpc$^{-2}$] & log $\Sigma$\rm{SFR} [$M_{\odot}$yr$^{-1}$kpc$^{-2}$] & log sSFR [yr$^{-1}$]  & $A_{\rm V}$ [mag] & $\beta_{\rm UV}$  \\ 
\hline  
$<0.16^{\prime\prime}$ & 9.91$_{-0.09}^{+0.11}$& -5.85$_{-36.21}^{+5.98}$& -15.76$_{-36.12}^{+5.86}$&  0.8$_{-0.4}^{+0.4}$ & 0.18$\pm$0.48 \\
$0.16^{\prime\prime}-0.35^{\prime\prime}$ & 9.52$_{-0.10}^{+0.07}$ & 0.73$_{-0.15}^{+0.14}$& -8.78$_{-0.06}^{+0.07}$&  1.7$_{-0.2}^{+0.2}$ & -0.38$\pm$0.30\\
$0.35^{\prime\prime}-0.5^{\prime\prime}$ & 8.94$_{-0.08}^{+0.10}$& 0.14$_{-0.14}^{+0.15}$& -8.80$_{-0.06}^{+0.05}$&  1.2$_{-0.2}^{+0.2}$ & -0.97$\pm$0.20\\
$0.5^{\prime\prime}-0.7^{\prime\prime}$ & 8.65$_{-0.09}^{+0.07}$& -0.13$_{-0.15}^{+0.16}$& -8.78$_{-0.06}^{+0.09}$&  1.0$_{-0.2}^{+0.2}$ & -1.18$\pm$0.19\\

\hline

\end{tabular}
\label{table1} 
\end{threeparttable} 
\end{table*}

\section{Physical properties of Zhúlóng}\label{Sec: measure}

\subsection{Appearance of bulge, disk, and spiral structures}\label{Sec: morphology}

We investigate the morphology of Zhúlóng with three deep JWST long-wavelength filters, F277W, F356W, and F444W, and their inverse-variance weighted stacked image. 
We fit our source with \texttt{PySersic} \citep{Pasha2023JOSS} using two different strategies: 1) a single S\'{e}rsic profile and 2) double S\'{e}rsic profiles. In both cases, we use PSFs for fitting which are derived from the WebbPSF software \citep{Perrin2014} and rotated to match the position angle of the observations. The single S\'{e}rsic fit returns the best-fit half-light radius ($R_{\rm e}$), which is used to quantify the galaxy size. The double S\'{e}rsic fit helps to distinguish between the contributions of the bulge and disk. In addition, comparing the best-fit results from single and double S\'{e}rsic profiles allows for an inspection of the underlying bulge in the core.

The fitting process is performed on $4^{\prime\prime}\times 4^{\prime\prime}$ background-subtracted cutouts centered on Zhúlóng. The bright neighboring foreground galaxy at $z\sim1.59$ is masked during the fitting. Additionally, we conduct tests comparing the results when masking the spiral arms, versus not masking them. We find that no matter whether masking them or not during the fitting process, for both single and double S\'{e}rsic fits, the best-fit results ($R_{\rm e}$, S\'{e}rsic index $n$, minor-to-major axis ratio $b/a$) remain consistent within the $1\sigma$ error. In this analysis, we focus on the results where the spiral arms are also masked during the fitting, as this approach enhances the visibility of the spiral arm structure in the residual plots, allowing us to more distinctly separate the disk, spiral arms, and the underlying bulge structure.

Fig.~\ref{morphology_plot} shows the best-fit results of the stacked image. From the residual maps, Zhúlóng shows evidence for a grand-design spiral arm structure, with high contrast primary arms extending from the bulge pointing north and south \citep[e.g.][]{Dobbs2014, Sellwood2022}. The presence of the spiral arms is also indicated in the surface brightness profiles, which show the discrepancy between the observed points and the best-fit model. The single S\'{e}rsic fit yields $R_{\rm e}=2.9 \pm 0.1$ kpc, suggesting that Zhúlóng has a large stellar size. The large $n=4.9\pm0.1$ reveals that the stellar light is more concentrated at the galaxy center, indicating the potential existence of a bulge structure. Meanwhile, the significant inhomogeneity in the core of the residual map also suggests the need for an additional bulge component in the fitting process, which is mitigated in the residual map of the double S\'{e}rsic fit. On the other hand, the double S\'{e}rsic fit yields evidence for both bulge + disk structures, with spiral arms shown in the residual map. The bulge component has $n=3.7\pm0.2$, close to the shape of classical bulges in the local Universe ($n\sim4$), suggesting the existence of a well-evolved bulge even in the first Gyr after the Big Bang. The bulge-to-total (light) ratio is $B/T=0.44$. The disk component has $n=1.2\pm0.1$, $R_{\rm e}=3.7\pm0.1$ kpc, and $b/a=0.99 \pm 0.01$, indicating the existence of a large face-on exponential stellar disk. In addition, we note that both of the residual maps show clear evidence of stellar clumps along the spiral arms, as observed in galaxies at cosmic noon \citep[e.g.,][]{Rujopakarn2016,Claeyssens2023,Claeyssens2024} and as also found in numerical simulations \citep[e.g.,][]{Tamburello2015,Mandelker2017,FenschBournaud2021,Ceverino2023,Renaud2024}. Furthermore, we also perform the morphological analysis for individual images, and the results remain the same (see Appendix~\ref{fit}). 

Overall, from the morphological analysis, it appears that the fit may be more in favor of the double S\'{e}rsic model. However, this alone does not constitute strong enough evidence to confirm the existence of a bulge and disk in Zhúlóng. As we will show in Sect.~\ref{Sec: annulus}, spatially resolved SED modeling reveals significant differences in the stellar populations between the inner and outer regions of the galaxy. The combination of the morphological analysis and the SED evidence suggests that the galaxy does indeed contain separate inner and outer structural components. 

Given the extended nature of Zhúlóng, and the amount of flux present in spiral arms that are not well-accounted for by either S\'{e}rsic model, we also measure the size that encloses 80\% of the light empirically, using the azimuthally-averaged light profile from the stacked image without any assumption on the parametric shape of the light distribution. Using this method, we find that these radii are larger than inferred assuming the parametric S\'{e}rsic (or double S\'{e}rsic) model. We measure R$_{80}$=9.5 kpc for the radii enclosing 80\% of the light. R$_{80}$ is extremely large compared to its half-light radius (while better reflecting the overall extent of stellar light), reflecting the fact that the spatial extent of the stars in the galaxy reaches more than 19 kpc in diameter.

\subsection{A quiescent-like bulge and star-forming disk}\label{Sec: annulus}

We further analyze different annular regions to explore any variation in the stellar populations across Zhúlóng. Physical properties are estimated by fitting the UV-to-NIR SED from JWST and HST photometry using \texttt{Bagpipes} \citep{Carnall2018}, with the $z_{\rm phot}$ fixed to that measured by \texttt{eazy}. We consider two different star formation histories (SFHs): a double-power-law and a delayed SFH, using \cite{Bruzual2003} stellar population models and the \cite{Calzetti2000} reddening law. We adopt a broad metallicity range of 0.1 to 2.5 $Z_{\odot}$, dust attenuation to the rest-frame V band ($A_{\rm V}$) values from 0 to 5 magnitudes, and ionization parameters log$U$ ranging from -4 to -2. For the double-power-law SFH, we adopt a falling slope $\alpha$ = [0.01, 1000], a rising slope $\beta$ = [0.01, 1000], and a turnover time $\tau$ = [0.01, 5] Gyr. For the delayed SFH, we apply broad uniform priors with age (i.e., the time since star formation began) ranging from 10 Myr to 1.2 Gyr and a logarithmic $\tau$ (timescale of decrease) ranging from 10 Myr to 10 Gyr. We obtain the main physical properties of $M_{\star}$, $A_{\rm V}$, and SFR, consistent within $1\sigma$ between the results assuming the two different SFHs. As an example, the best-fit SED results assuming a double-power-law SFH are shown in Fig.~\ref{fig3} and Table \ref{table1}.

In Fig.~\ref{fig3}, from the central core to the outer region, the best-fit SED changes significantly from quiescent to star-forming populations. The core is dominated by old stellar populations with a red color ($F150W-F444W=3.1$ ABmag). The best-fit SED exhibits a strong Balmer/4000 \AA~break, very low $\log$(sSFR/yr$^{-1})=-15.76_{-36.12}^{+5.86}$, and a lack of emission lines. The observed colors are consistent with NIRCam-based photometric selections for $z>3$ quiescent galaxies \citep{Long2024}. The rest-frame U-V and V-J colors as measured from \texttt{eazy} are 1.4 and 1.0 mag, respectively, consistent with a quiescent/post-starburst fading stellar population \citep{williams2009}. The restframe colors are also consistent with extensions of UVJ color criteria for quiescent galaxies that have been adapted to $z>3-6$ to accommodate younger passively evolving galaxies  \citep{AntwiDanso2023}. In contrast, the disk measured in an annulus between 0.5$^{\prime\prime} - 0.7^{\prime\prime}$ is dominated by younger stellar populations with a relatively bluer color ($F150W-F444W=2.0$ ABmag) and, unlike the inner region, allows the presence of weak emission lines in the SED. 

Furthermore, among all annular regions, the central core ($r<0.16^{\prime\prime}$) has the highest stellar mass surface density, with log($\Sigma M_{\rm \star}/M_{\odot}$ kpc$^{-2}) = 9.91_{-0.09}^{+0.11}$ (see Table \ref{table1}), revealing the build-up of a dense stellar bulge. The absence of star formation activity in the core is consistent with the expectations of inside-out galaxy growth and quenching \citep[e.g.][]{Tacchella2015}. 

Finally, we also calculate the stellar mass and stellar mass surface density of the bulge. Assuming the core is dominated by the bulge, with the mass-to-light ratio in the core and the best-fit bulge light profile (Fig.~\ref{morphology_plot}), we derive $\log (M_{\star, \rm bulge}/M_{\odot})=10.71_{-0.09}^{+0.10}$ and log($\Sigma M_{\rm \star, \rm bulge}/M_{\odot}$ kpc$^{-2}) = 10.00_{-0.11}^{+0.09}$. Given the total mass of Zhúlóng (see Sect.~\ref{Sec: mass}), its bulge-to-total mass ratio is $B/T=0.5$.

\subsection{Ultra-massive spiral galaxy with accelerated formation }\label{Sec: mass}
Here we focus on the integrated properties of Zhúlóng. The total $M_{\star}$, $A_{\rm V}$, and SFR are derived by fitting the UV-to-FIR SED from the JWST+HST+ALMA photometry, using \texttt{Bagpipes}. Similar to Sect.~\ref{Sec: annulus}, we use the \cite{Calzetti2000} reddening law, the \cite{Bruzual2003} stellar population models, and the \cite{Draine2007} dust emission model (with energy balance), assuming different SFH (double-power-law SFH and delayed SFH) and free metallicity. The derived main properties with delayed SFH are listed in Table \ref{table1}. To test the result, we also perform SED fitting with \texttt{CIGALE} \citep{Boquien2019}, which produces very consistent values within the errors. The total stellar mass is also consistent with the sum of the individual masses obtained from the different annuli ($\log (M_{\star}/M_{\odot})=10.96_{-0.06}^{+0.05}$; see Sect.~\ref{Sec: annulus}). In general, we find that Zhúlóng is extremely massive with $\log (M_{\star}/M_{\odot})=11.03_{-0.08}^{+0.10}$.

\begin{figure}
\centering
\includegraphics[width=9cm]{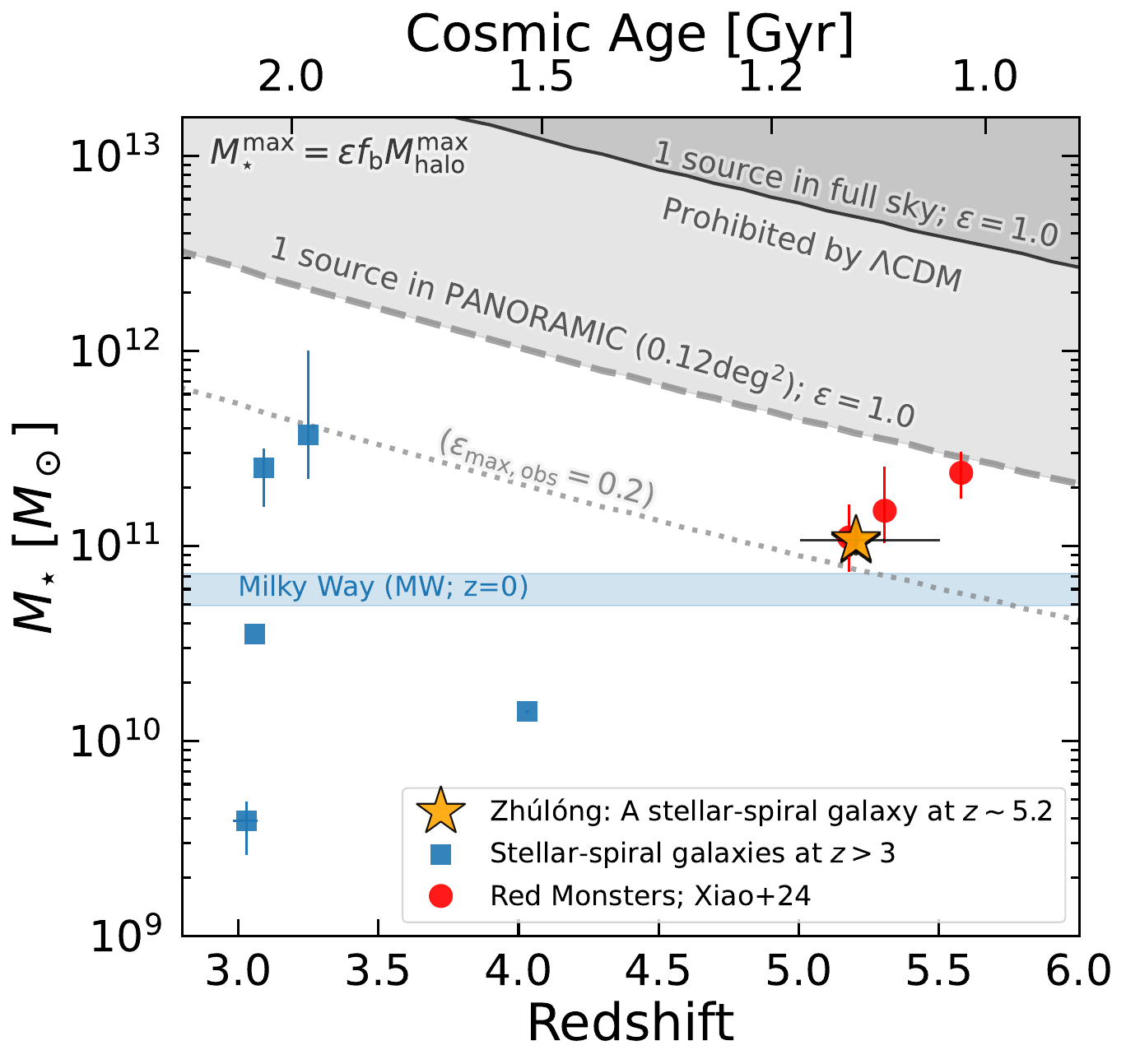}
\caption{\textbf{Stellar mass of Zhúlóng (orange star) compared to other representative galaxies and the model expectations.} The red-filled circles show three spectroscopically-confirmed ultra-massive galaxies \citep[so-called "red monsters";][]{Xiao2024}. The blue squares show $z>3$ galaxies reported to have clear stellar spiral structures \citep{jain2024, Wangweichen2024, Umehata2024, Costantin2023, Wu2023}. Note that there is no error bar of $M_{\star}$ given in \cite{Wu2023}. The grey dashed and dotted lines indicate the maximum stellar mass calculated from the maximum halo mass ($M_\mathrm{halo}^\mathrm{max}$) given the PANORAMIC survey volume, based on $M_{\star}^\mathrm{max} = \epsilon f_{\rm b} M_\mathrm{halo}^\mathrm{max}$, the cosmic baryon fraction $f_{\rm b} = \Omega_{\rm b}/\Omega_{\rm m} = 0.158$, and assuming a baryon-to-star conversion efficiency of $\epsilon=1$ and 0.2, respectively. As a reference, we also show $\epsilon=1$ with the full sky coverage as the black line. The grey regions indicate the stellar mass prohibited by the standard $\Lambda$CDM cosmology.   
      }
         \label{fig2}
\end{figure}

Such high stellar mass at $z\sim5.2$ indicates that Zhúlóng must have been forming stars very efficiently. In Fig.~\ref{fig2}, we compare its stellar mass to the maximum mass at which one would expect to find a galaxy within the PANORAMIC survey volume, given the prevailing halo mass function and cosmic baryon fraction \citep{Boylan-Kolchin2023, Lovell2023}. Instead of taking the small coverage of PANORAMIC in the COSMOS field only, to be conservative, we consider the total PANORAMIC survey area of $\sim$ 432 arcmin$^{2}$. Under this paradigm, we derive the most massive dark matter halo mass ($M_\mathrm{halo}^\mathrm{max}$) at different redshifts within the corresponding survey volume ($\Delta z=1$; e.g., $\sim1.2\times10^{6}$ Mpc$^3$ at $z=4.7-5.7$) according to the halo mass function. The derived $M_\mathrm{halo}^\mathrm{max}$ at $z=5.2$ is log($M_{\rm halo}/M_{\odot}$) $=12.38$. 

The maximum stellar mass is inferred from the maximum dark matter halo mass, based on $M_{\star}^\mathrm{max} = \epsilon f_{\rm b} M_\mathrm{halo}^\mathrm{max}$, with a cosmic baryon fraction $f_{\rm b} = \Omega_{\rm b}/\Omega_{\rm m} = 0.158$ \citep{Planck2020}, and the maximum theoretical efficiency ($\epsilon$) of converting baryons into stars. Here, we consider two possible cases of $\epsilon$ in the PANORAMIC survey volume, as shown in Fig.~\ref{fig2}: the highest efficiency from observation-based phenomenological modeling, such as abundance matching and halo occupation distribution models \citep[$\epsilon_{\rm max,obs} = 0.2$; dotted line;][]{Moster2013, Moster2018, Tacchella2018, Pillepich2018, Wechsler2018, Shuntov2022} and the maximum efficiency logically allowed ($\epsilon=1$; dashed line). As a reference, we also overplot the most extreme case of $\epsilon=1$ in the full sky (solid line).

By comparing the stellar mass of Zhúlóng with the predictions in Fig.~\ref{fig2}, it is clear that Zhúlóng requires an extremely efficient conversion of available baryons to stars of $\epsilon\sim0.3$ -- about 1.5 times the highest efficiency observed at lower redshift ($\epsilon_{\rm max,obs} \simeq 0.2$). The ultra-massive properties and high efficiency make Zhúlóng another extreme source in addition to recently discovered massive galaxies \citep[e.g.,][]{Xiao2024, glazebrook2017,Glazebrook2024,Carnall2023Natur, Weibel2024b, deGraaff2024}. In Fig.~\ref{fig2}, we show a direct comparison of Zhúlóng and three red monster galaxies at $z_{\rm spec}\sim5-6$ \citep[red points; i.e., S1, S2, and S3 in][]{Xiao2024} as their efficiency is calculated within the similar survey volume.
Together, these findings demonstrate the existence of ultra-massive galaxies that challenge our galaxy assembly models in the first billion years of the Universe \citep{White1978, Boylan-Kolchin2023, Labbe2023a}. 

In addition, in Fig.~\ref{fig2}, we compile all currently known examples from the literature of galaxies whose stellar structures are consistent with spiral arms at $z>3$, the highest redshift at which spiral structure has been detected in stars (rather than just gas) to date \citep[][]{jain2024, Wangweichen2024, Umehata2024, Costantin2023, Wu2023}. We note that while some other galaxies have been reported to show potential spiral-like structures in JWST images, they are not included in our figure due to the lack of comparable analyses and stellar mass estimates \citep[e.g.,][]{Ferreira2023}. Thus, we caution that there may be more spiral galaxies at $z>3$ than those currently identified, and a systematic study of high-redshift spiral galaxies is necessary to confirm this. In addition, sources showing only stellar disks but without clear spiral arms are also not in our plot \citep[e.g.,][]{Yan2024}. Compared with the literature-reported spiral galaxies, Zhúlóng appears to be the most distant spiral galaxy discovered so far. 

\subsection{The location on the main sequence}

\begin{figure}
\centering
\includegraphics[width=8cm]{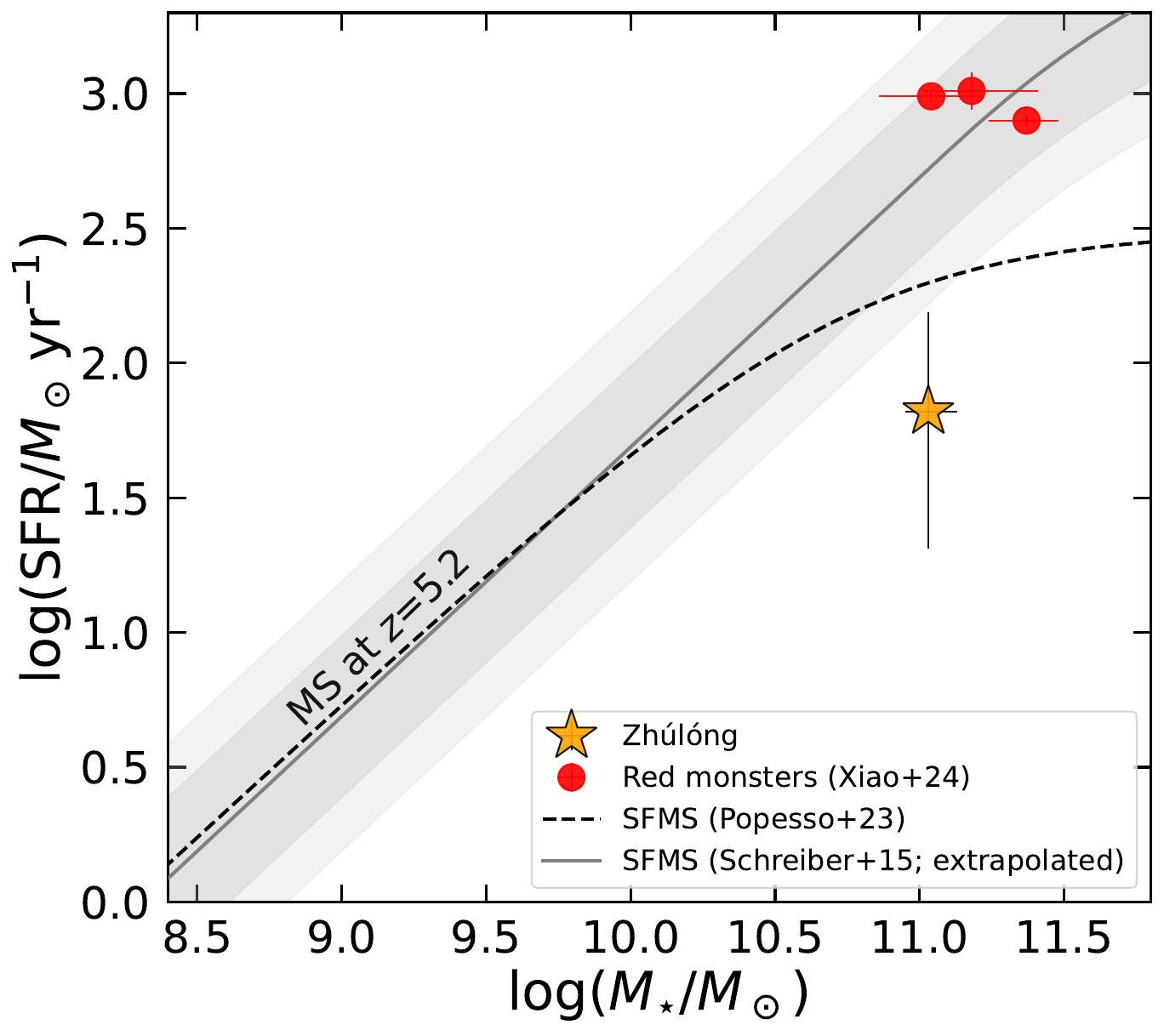}
\caption{\textbf{Location of Zhúlóng (orange star) compared to the SFMS in the SFR-$M_{\star}$ plane.}  The red-filled circles show three spectroscopically-confirmed ultra-massive galaxies \citep{Xiao2024}. Error bars correspond to 1$\sigma$ uncertainties.  The extrapolated \cite{Schreiber2015} SFMS at $z=5.2$, 1$\sigma$ scatter ($\sim$0.3 dex), and a more extended typical scatter of $0.5$ dex are highlighted with a grey line, a dark grey shaded area, and a light grey shaded area, respectively. We also plot the \cite{Popesso2023} SFMS at $z=5.2$  with a dashed line.
      }
         \label{fig_ms}
\end{figure}

Given the red, quiescent-like bulge, we investigate the location of our galaxy with respect to the star-formation main sequence (SFMS) at $z\sim5$. While previous ultra-massive galaxies are found to be in a rapid star-formation phase, Zhúlóng's star-formation rate is significantly low. For example, the three sources from \citet{Xiao2024} S1, S2, and S3 are ultra-massive, dusty star-forming, sub-millimeter galaxies. They have SFRs of $\sim$1000 $M_{\odot}$yr$^{-1}$, large dust contents ($A_{\rm V}>3$ mag and detected by SCUBA2 observations), half-light radii of $R_{\rm e, F444W}\sim 1-2$ kpc. Due to heavy dust obscuration, they are barely detectable in the optical wavelength bands, and hence they are also called optically dark galaxies \cite[or HST-dark galaxies; e.g.,][]{wangtao2019, Williams2019, Xiao2023, Williams2024_RED}.   By contrast, as shown in Fig.~\ref{fig_ms}, the SFR of Zhúlóng is 66$_{-46}^{+89}$ $M_{\odot}$yr$^{-1}$ (also in Table \ref{table1}), more than 10 times lower than S1, S2, and S3. It is not detected in the deep ALMA observations at 1.2mm (3$\sigma$ upper limit of 0.6 mJy beam$^{-1}$) and other FIR observations, consistent with its low measured SFR (see Sect.~\ref{alma}). With these properties, the galaxy is about 1 dex below the main sequence as characterized by \citet{Schreiber2015}, and 0.5 dex below the relation of \cite{Popesso2023}.

\section{Implications of the emergence of an ultra-massive grand-design spiral galaxy at 1 Gyr}\label{Sec: disc}

\subsection{An outlier among massive galaxies at $z\sim5$ }

As discussed in Section \ref{Sec: morphology}, the galaxy is large, extending over 19 kpc (based on the size enclosing 80\% of the light). We also measure its half-light size based on 1- and 2-component S\'{e}rsic profiles, resulting in $\sim3$ and 3.9 kpc (for each model, respectively). By all metrics, Zhúlóng is large relative to comparable galaxies at similar redshifts. The typical average half-light size at $z\sim5$ at $\log M_*/M_\odot\sim10.7$ from recent studies range between 0.4-2 kpc \citep[e.g.,][]{Allen2024, Ito2024, Ormerod2024, Varadaraj2024, Ward2024}, as estimated based on single-S\'{e}rsic models. These findings are in line with the expectation that galaxy size scales with dark matter halo size, which decreases with increasing redshift \citep[e.g.][]{mo1998}. 

Unfortunately, JWST-derived mass-size relations, to date, are based almost entirely on galaxies one order of magnitude lower in mass, complicating direct comparison with this source. Compared to an extrapolation to the size-mass relation of SFGs at similar redshift, the half-light radius of Zhúlóng (Re$\sim$3 kpc; single-S\'{e}rsic) does fall within the extrapolated scatter from lower masses \citep[e.g.][]{Allen2024, Miller2024}.

However, in light of its low integrated sSFR  ($\log$ sSFR$\sim-9.2$ yr$^{-1}$; $>$0.5 dex below the SFMS at these redshifts; see Fig.~\ref{fig_ms}), such a large size is quite remarkable. Its red rest-frame U-V vs V-J colors in the core (see Section \ref{Sec: annulus}) are consistent with quiescent systems, perhaps making a comparison to that population more appropriate. Typically, UVJ-red galaxies follow a steeper mass-size relation leading to much smaller sizes at fixed stellar mass than SFGs \citep[e.g.,][]{vanderwel2014, Ito2024, Ji2024b}. Quiescent galaxies with similarly high mass ($\log M_*/M_\odot>$10.3-11) at comparable redshifts \citep[$z>4-7$; e.g.,][]{Carnall2023Natur, deGraaff2024, Weibel2024b, Ji2024c, Ito2024}, have Re $\sim 0.4-0.7$ kpc, a factor of $>4\times$ smaller than what we find here (using our single-S\'{e}rsic measurement for Zhúlóng, for consistency with other studies). Unfortunately, sample sizes of comparable galaxies at $\log M_*/M_\odot\sim11$ at $z\sim5$ remain small, regardless of sSFR, and preclude a robust conclusion in comparison to the population scatter.

Thus, we also compare to the R$_{80}$ sizes of very massive $\log M_*/M_\odot\sim11$ galaxies (across all sSFR) from larger-area HST surveys at $z\lesssim3$ \citep{Mowla2019}. We find that R$_{80}$ for this galaxy is still 50$\%$ larger than the average R$_{80}$ among comparable mass galaxies that appear more than 1 Gyr later \citep[$z\sim2.75$;][]{Mowla2019b}. While a definitive comparison to typical co-eval massive galaxies is not possible without a more detailed analysis and data on larger JWST-based samples, collectively these first comparisons imply that Zhúlóng could be a notable outlier among massive galaxies, regardless of its location relative to the MS at $z>4$.

While it is a likely outlier due it its large size, its properties as a massive spiral galaxy at $z\sim 5.2$ are also unexpected. The fraction of visibly spiral galaxies is already known to drop down to 4\% at $z\lesssim 3$, in part due to surface brightness dimming that impacts visibility as $(1+z)^4$ \citep[][]{Kuhn2024}, but remain rare even at cosmic noon \citep{Law2012,Yuan2017}. The entire spiral disk is also quite red; even at $z\lesssim2$, where spirals are more prevalent \citep{Conselice2014}, those with similarly red color \citep[restframe 0.3-0.5$\mu$m $\gtrsim$2 ABmag;][]{Fudamoto2022} are rare, only making up 2$\%$ of the galaxy population \citep{Shimakawa2022}. The rarity of similar sources later in cosmic time makes the existence of Zhúlóng at $z\sim5.2$ remarkable.

\subsection{A Milky Way analog with 1 Gyr formation time}

While a potential outlier at $z\sim5$, this galaxy is instead remarkably comparable to massive evolved spirals like the MW at $z\sim0$. MW-like spirals are characterized by old bulges (thought to form early in the universe) but whose extended disks are thought to have built up slowly over the course of 10 Gyr \citep{vanDokkum2013, RixBovy2013}. For context, current estimates indicate that the MW has a comparably massive core (bulge + bar) to Zhúlóng with $\log M_*/M_\odot \sim10$, total stellar mass of $\log M_*/M_\odot\sim10.8$ (even slightly lower than Zhúlóng), with comparable sSFR, consistent with our upper limit \citep[$\sim 10^{-10.6}$ yr$^{-1}$;][]{Licquia2015}. Various estimates for the half-light radius of the MW suggest a scale length for the thin disk of $R_{e}=2.6\pm0.5$ kpc \citep[][and references within, based on independent analyses ranging between $1.8-6$ kpc]{BlandHawthorn2016}. The full extent of its stellar disk (diameter $\sim19$ kpc) is already of comparable size to mature MW-like spirals seen at $z\sim0$ \citep[$\sim8-12$ kpc in diameter; e.g.][]{BlandHawthorn2016, Lian2024}. 

Unlike typical co-eval galaxies, Zhúlóng has instead developed comparable structures (bulge mass and density, disk and spiral size, spatially segregated stellar populations) and comparable stellar mass, 10$\times$ faster than MW-like galaxies at $z\sim0$, demonstrating that such an advanced evolutionary stage in galaxy evolution can be reached in only 1 Gyr time. Moreover, in $\Lambda$CDM, the Universe is expected to be 6$^3$ denser at z$\sim$5 compared to today's Universe, this makes it astonishing that Zhúlóng appears with a well-organized morphology with a similar z$\sim0$ MW's size, despite this high merger rate in the early Universe \citep[e.g.][]{Duan2024}. How a morphologically mature galaxy that resembles nearby massive spirals can form in this environment remains an open question, but the discovery of this source is a first step and provides an important constraint on galaxy formation models.

\subsection{Implications for efficient growth of massive galaxies}

The ultra-massive nature of Zhúlóng indicates an efficient mass build-up and a high star formation efficiency (see Fig.~\ref{fig2}) in the past. Multiple minor mergers and/or major mergers may have driven this rapid growth. In addition, the presence of clumps also supports efficient growth;  fragmentation of gas-rich disks into star-forming clumps may increase the star formation efficiency, helping galaxies grow more quickly to reach such masses \citep[e.g.,][]{Dessauges-Zavadsky2019, Dessauges-Zavadsky2023, Faisst2024, Fujimoto2024}. 

Although the galaxies described below have all been reported with high star formation efficiencies, the structure in the stellar populations of Zhúlóng clearly distinguishes it from other ultra-massive dusty star-forming galaxies \citep[e.g.,][]{Xiao2024} and early-forming massive quiescent galaxies \citep[e.g.,][]{glazebrook2017,Glazebrook2024,Carnall2023Natur, Weibel2024b, deGraaff2024}. It is clear that Zhúlóng formed through an inside-out pathway (see Section \ref{Sec: annulus}), and only a handful of other galaxies at $z>4$ have been identified to have this stellar structure \citep[e.g.,][]{Baker2024, Ji2024c, Nelson2024}. 

Zhúlóng has now demonstrated that there must be broad diversity in morphology among ultra-massive galaxies. While an answer about the origin of the spiral structure awaits further spectroscopic information, it now seems clear that at least some ultra-massive galaxies exhibit large disks (in contrast to the other known examples, which are nearly all compact). This may reflect a diversity of formation pathways that must be factored into any theories about the physical processes that drive their early formation at high redshift. Larger sample sizes will be required to better understand the morphological diversity of ultra-massive galaxies, which will be enabled in the future by the continued accumulation of wide-area JWST imaging.

\section{Conclusions}\label{Conclusions}
In this paper, we present the serendipitous discovery of the most distant ultra-massive spiral galaxy candidate identified in the pure parallel imaging survey PANORAMIC (in the COSMOS field). It has $z_{\rm phot} = 5.2^{+0.3}_{-0.2}$ well-constrained by a strong Balmer/4000 \AA ~break and consistent with the neighboring star-forming clump (Fig.~\ref{fig1}). After combining the JWST/NIRCam, HST, and ALMA data, we find the following extremely intriguing properties:

\noindent
\textbf{1. Quiescent bulge + large star-forming disk + spiral arms:} 
From the morphological analysis, Zhúlóng perfectly exhibits a classical bulge ($n=3.7\pm0.2$ and $R_{\rm e}=0.9\pm0.1$ kpc), a large face-on stellar disk ($n=1.2\pm0.1$, $R_{\rm e}=3.7\pm0.1$ kpc, and $b/a=0.99 \pm 0.01$) extending over 19 kpc, with spiral arms embedded in (Fig.~\ref{morphology_plot}). The SED analysis on different annular regions shows Zhúlóng has a quiescent-like core and a star-forming stellar disk (Fig.~\ref{fig3}). Compared to the stellar disk, the center core is $i$) red ($F150W-F444W=3.1$ mag); $ii$) has among the highest stellar mass surface densities measured among quiescent galaxies (log($\Sigma M_{\rm \star}/M_{\odot}$ kpc$^{-2}) = 9.91_{-0.09}^{+0.11}$; see Table \ref{table1}) 
-- revealing the build-up of the stellar bulge; and $iii$) and is quiescent -- indicating an inside-out galaxy growth (or quenching). For the bulge, we derive $\log (M_{\star, \rm bulge}/M_{\odot})=10.71_{-0.09}^{+0.10}$, log($\Sigma M_{\rm \star, \rm bulge}/M_{\odot}$ kpc$^{-2}) = 10.00_{-0.11}^{+0.09}$, and $B/T=0.5$.

\noindent
\textbf{2. Extremely massive:} Zhúlóng is extremely massive ($\log (M_{\star}/M_{\odot})=11.03_{-0.08}^{+0.10}$; Fig.\ref{fig2}), adding to a growing population of ultra-massive galaxies discovered in the first billion years. Assuming it is located in the most massive dark matter halo expected in the whole PANORAMIC survey volume, we infer that Zhúlóng must have been forming stars very efficiently, with the baryons-to-stars conversion efficiency of $\epsilon\sim0.3$ -- about 1.5 times higher than even the most efficient galaxies at later epochs.

\noindent
\textbf{3. Low star formation activity:} Although the disk is still forming stars, in total, Zhúlóng has SFR $=66_{-46}^{+89}$ $M_{\odot}$yr$^{-1}$ (Table \ref{table1} and Fig.\ref{fig_ms}),  $\gtrsim$0.5 dex below the SFMS at $z\sim5.2$ \citep{Schreiber2015,Popesso2023}. The relatively modest SFR of Zhúlóng indicates that it is in the transformation phase from star-forming to quiescence. 

\noindent
\textbf{4. Most distant spiral galaxy discovered so far:} Compared to other recently reported stellar spiral galaxies at $z\sim3$, Zhúlóng appears to be the most distant spiral galaxy discovered so far. While the origin of the spiral structure is unknown, it represents an important example that demonstrates that a MW-like galaxy can evolve earlier in the universe than previously thought ($10\times$ faster than locally, $<1$Gyr after the Big Bang), while efficiently building enormous stellar mass. 

Altogether, Zhúlóng at $z_{\rm phot}\sim5.2$ is potentially the highest-redshift, red, ultra-massive, grand-design spiral galaxy with a large stellar size. Overall, the galaxy sits below the main sequence, suggesting the star formation activity is ramping down. 
It contains a quiescent bulge + large star-forming disk + spiral arms. For a galaxy at $z>5$, having any of these relevant properties would be enough to make the galaxy special. The presence of all of these properties makes Zhúlóng very exceptional, indicating the rapid formation and morphological evolution of massive galaxies in the early universe. 
With future JWST IFU observations, combined with higher-resolution and deeper ALMA data, this galaxy will provide a unique window to unveil what triggers the rapid mass assembly together with the early disk formation.

\begin{acknowledgements}
We are very grateful to the anonymous referee for instructive comments, which helped to improve the overall quality and strengthened the analyses of this work. 
We thank Boris Sindhu Kalita, Francoise Combes, and Maximilien Franco for valuable discussions and suggestions that improved this paper.
This work is based in part on observations made with the NASA/ESA/CSA James Webb Space Telescope. The data were obtained from the Mikulski Archive for Space Telescopes at the Space Telescope Science Institute, which is operated by the Association of Universities for Research in Astronomy, Inc., under NASA contract NAS 5-03127 for JWST. These observations are associated with program 2514. Support for program JWST-GO-2514 was provided by NASA through a grant from the Space Telescope Science Institute, which is operated by the Association of Universities for Research in Astronomy, Inc., under NASA contract NAS 5-03127. 
This paper makes use of the following ALMA data: ADS/JAO.ALMA\#2023.1.00180.L. ALMA is a partnership of ESO (representing its member states), NSF (USA) and NINS (Japan), together with NRC (Canada), NSTC and ASIAA (Taiwan), and KASI (Republic of Korea), in cooperation with the Republic of Chile. The Joint ALMA Observatory is operated by ESO, AUI/NRAO and NAOJ. The National Radio Astronomy Observatory is a facility of the National Science Foundation operated under cooperative agreement by Associated Universities, Inc.
This work has received funding from the Swiss State Secretariat for Education, Research and Innovation (SERI) under contract number MB22.00072, as well as from the Swiss National Science Foundation (SNSF) through project grant 200020\_207349. 
The work of CCW is supported by NOIRLab, which is managed by the Association of Universities for Research in Astronomy (AURA) under a cooperative agreement with the National Science Foundation.
The Cosmic Dawn Center (DAWN) is funded by the Danish National Research Foundation under grant DNRF140. 
PD acknowledges support from the NWO grant 016.VIDI.189.162 (``ODIN") and warmly thanks the European Commission's and University of Groningen's CO-FUND Rosalind Franklin program. 
AH acknowledges support by the VILLUM FONDEN under grant 37459. 
MVM acknowledges support from the National Science Foundation via AAG grant 2205519 and the Wisconsin Alumni Research Foundation via grant MSN251397. CCW, AC, ZJ and KEW acknowledge funding from program JWST-GO-2514 that was provided by NASA through a grant from the Space Telescope Science Institute, which is operated by the Association of Universities for Research in Astronomy, Inc., under NASA contract NAS 5-03127.
ZJ also acknowledges funding from JWST/NIRCam contract to the University of Arizona NAS5-02015. 
I.L. acknowledges support from Australian Research Council Future Fellowship FT220100798.
SL acknowledges support by the Science and Technology Facilities Council (STFC) and by the UKRI Frontier Research grant RISEandFALL. 
LB acknowledges funding from STFC through ST/T000473/1 and ST/X001040/1.
K.G. and T.N. acknowledge support from Australian Research Council Laureate Fellowship FL180100060.

\end{acknowledgements}

\bibliography{reference_update}{}
\bibliographystyle{aa}

%
%

\onecolumn
\begin{appendix}

\section{Robustness of the redshift solution for Zhúlóng.}\label{low-z} 

\begin{figure*}[h!]
\centering
\includegraphics[width=18cm]{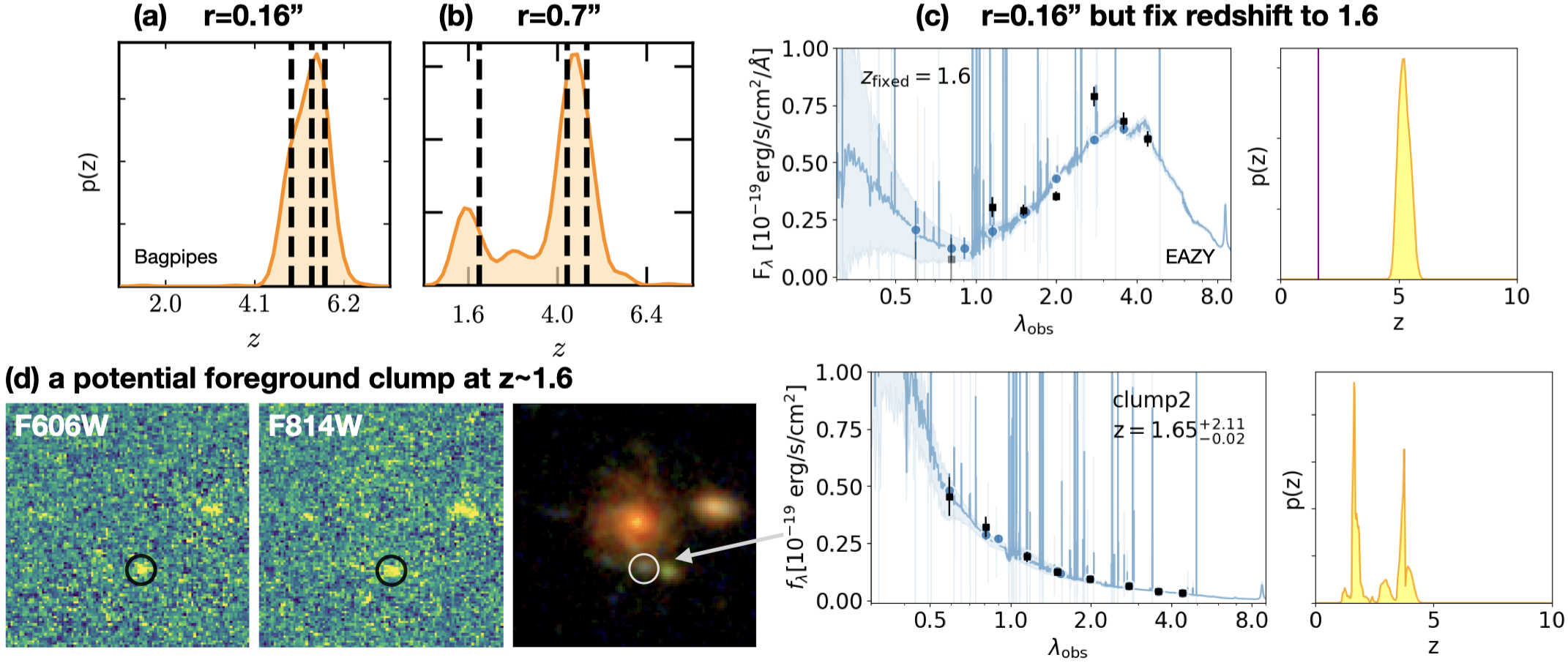}
\caption{(a) The photometric redshift likelihood distribution \( P(z) \) for the central core (0.16$^{\prime\prime}$ radius aperture), derived with \texttt{Bagpipes}, consistent with the results in Fig.~\ref{fig1}. 
(b) \( P(z) \) derived with \texttt{Bagpipes} using a 0.7$^{\prime\prime}$ radius aperture, showing a second redshift solution at \( z \sim 1.6 \). 
(c) The \texttt{EAZY} SED fitting results for the central core (0.16$^{\prime\prime}$ radius) with the redshift fixed at \( z = 1.6 \), indicating a poor fit with \( \chi^2 = 46 \). 
(d) A bright clump (clump 2, marked with a black circle), was detected at 5.5\( \sigma \) and 7.5\( \sigma \) in the F606W and F814W bands, respectively. The \texttt{EAZY} fit for clump 2 shows a broad redshift distribution with \( z_{\rm phot} = 1.65^{+2.11}_{-0.02} \), suggesting it may be a foreground object.
}
         \label{lowz}
\end{figure*}

In Fig.~\ref{fig1}, we show the narrow photometric redshift distribution \( P(z) \) of Zhúlóng, derived using \texttt{EAZY} for the galaxy core (0.16$^{\prime\prime}$ radius aperture). To ensure the robustness of our results, we also perform SED fitting with \texttt{Bagpipes}, allowing the redshift to vary from 1 to 10. We assume different SFHs, including delayed SFH and double-power-law SFH, and the same settings as described in Sect.~\ref{Sec: annulus}. The resulting \( P(z) \) shows a single strong peak, consistent with the \texttt{EAZY} results. Specifically, the photometric redshift is \( z_{\rm phot} = 5.42^{+0.31}_{-0.45} \) for delayed SFH (Fig.~\ref{lowz}-$a$) and \( z_{\rm phot} = 5.59^{+0.29}_{-0.38} \) for the double-power-law SFH.

However, when a larger aperture (e.g., 0.7$^{\prime\prime}$ radius) is used, a second redshift solution at \( z \sim 1.6 \) emerges (Fig.~\ref{lowz}-$b$). To further investigate this, we rerun the SED fitting for the galaxy core using \texttt{EAZY} with the redshift fixed to \( z = 1.6 \). The resulting fit, shown in Fig.~\ref{lowz}-$c$, has a large chi-square value (\( \chi^2 = 46 \)), indicating a very poor fit.

To explore the cause of this low-redshift solution with a large aperture, we further inspect the images and find a bright clump (clump 2, thereafter; Fig.~\ref{lowz}-$d$) at F606W and F814W from HST/ACS. Clump 2 is located south-east of the \( z_{\rm phot} = 5.15^{+0.04}_{-0.03} \) clump reported in Fig.~\ref{fig1}. The \texttt{EAZY} fitting results for clump 2 show a broad \( P(z) \) distribution with \( z_{\rm phot} = 1.65^{+2.11}_{-0.02} \), suggesting it may be a foreground clump. Thus, the second redshift peak in Fig.~\ref{lowz}-$b$ is likely due to contamination from clump 2.

Overall, we perform SED fitting on the galaxy core using both \texttt{EAZY} and \texttt{Bagpipes}, and both methods consistently indicate that Zhúlóng is a strong spiral galaxy candidate at $z>5$. However, we identify the potential presence of foreground structure(s) that could contaminate the redshift measurements of Zhúlóng when larger apertures are used. We note here that clump 2 does not affect our main results, as the global properties of Zhúlóng are derived using the total flux measured within a 0.5$^{\prime\prime}$ radius aperture (with aperture corrections applied), with clump 2 not included. While clump 2 is included in the SED fitting for the annular radius of 0.5$^{\prime\prime}-0.7^{\prime\prime}$ and may influence the fitting results (see Fig.~\ref{fig3}), we cannot confirm whether it is a low-$z$ clump without spectroscopic observations and thus did not remove it from the analysis. In any case, the key findings of inside-out galaxy growth and a quiescent-like galaxy core embedded in a star-forming stellar disk remain consistent even when considering only the remaining annuli. In addition, it does not affect our identification of spiral features. To confirm whether Zhúlóng is indeed a distant spiral galaxy at \( z > 5 \), follow-up JWST NIRSpec/IFU observations are necessary.

\clearpage
\section{Morphological modeling of Zhúlóng.}\label{fit}

\begin{figure*}[h!]
\centering
\includegraphics[width=13cm]{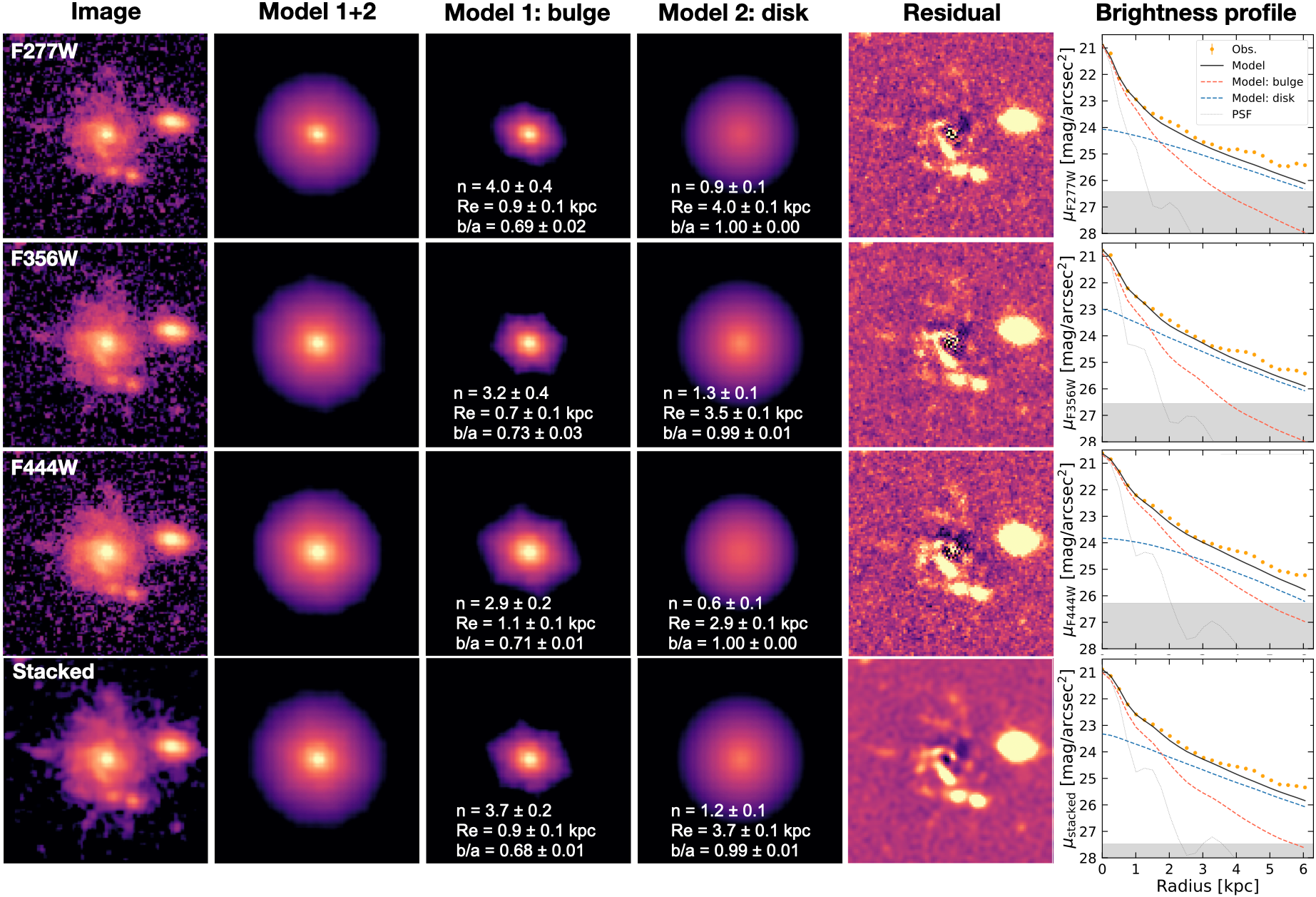}
\caption{Similar to Fig.~\ref{morphology_plot}, but showing results for the F277W, F356W, F444W, and stacked images, respectively. Each is modeled using double S\'{e}rsic profiles to decompose the bulge and disk.
      }
         \label{morphology_plot_2sersic}
\end{figure*}

\begin{figure*}[h!]
\centering
\includegraphics[width=10cm]{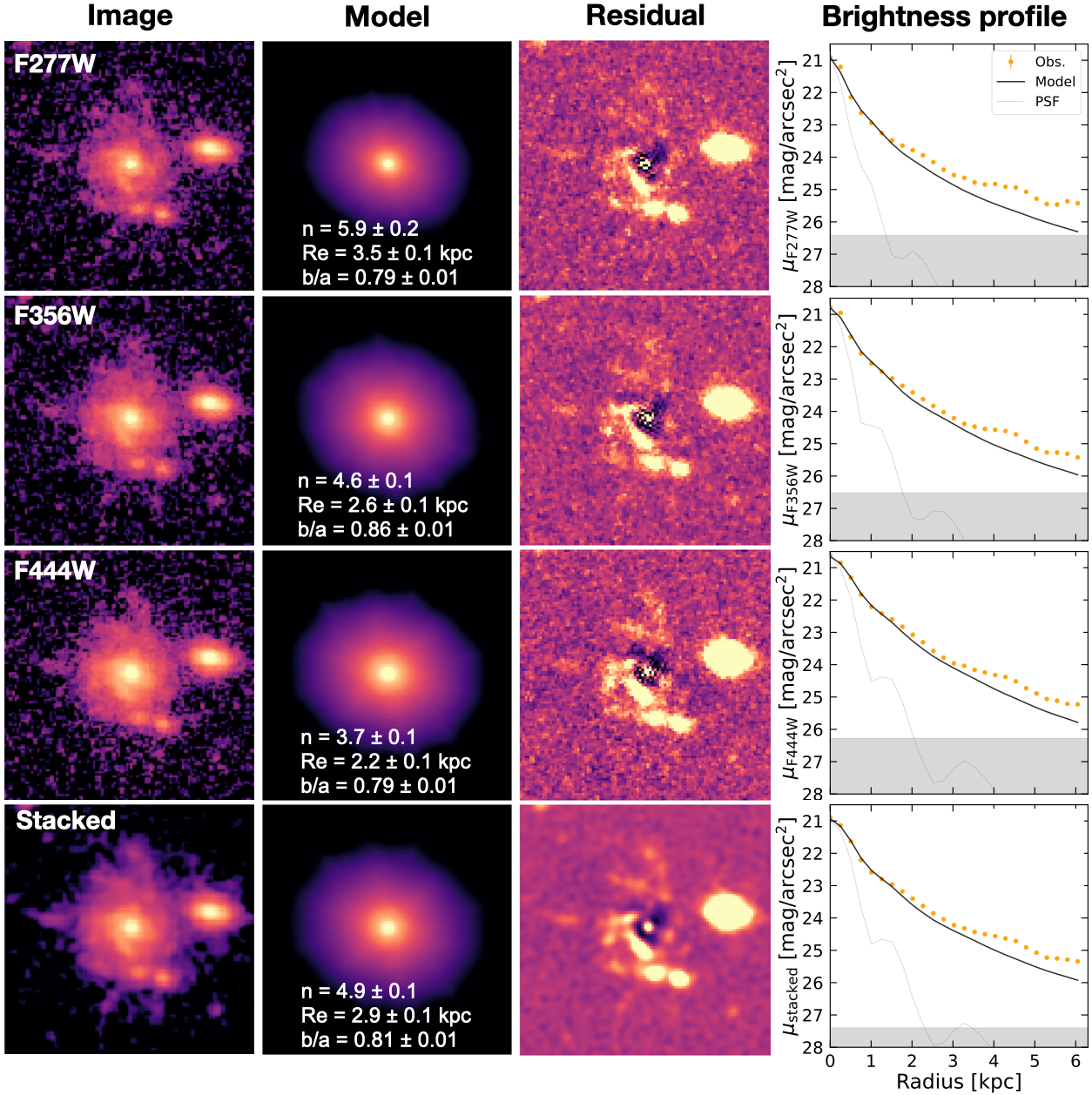}
\caption{Similar to Fig.~\ref{morphology_plot}, but showing results for the F277W, F356W, F444W, and stacked images, respectively. Each is modeled using a single S\'{e}rsic profile.
      }
         \label{morphology_plot_1sersic}
\end{figure*}

\end{appendix}

\end{document}